\documentclass{article}
\usepackage{PRIMEarxiv}

\usepackage[utf8]{inputenc} 
\usepackage[T1]{fontenc}
\usepackage{hyperref}
\usepackage{booktabs}    
\usepackage{multirow}    
\usepackage{adjustbox}   
\usepackage[table,dvipsnames]{xcolor}
\usepackage{colortbl}    
\usepackage{caption}     
\usepackage[most]{tcolorbox}
\usepackage{amssymb} 
\tcbuselibrary{skins}

\usepackage{array}       
\usepackage{graphicx}    
\usepackage{bold-extra}
\usepackage{marvosym}
\usepackage{orcidlink}
\usepackage{colortbl}
\usepackage{array}
\usepackage[ruled,vlined]{algorithm2e}
\definecolor{evo}{HTML}{f4d1de}
\usepackage{eurosym}

\usepackage{graphicx}
\usepackage{listings}
\usepackage{lipsum}

\lstdefinestyle{mintedclone}{
  language=C++,
  basicstyle=\ttfamily\scriptsize,
  numbers=left,
  numberstyle=\tiny\color{gray},
  frame=lines,
  rulecolor=\color{gray!60},
  keywordstyle=\color{blue!70!black},
  commentstyle=\color{gray!70},
  stringstyle=\color{orange!80!black},
  showstringspaces=false,
  tabsize=2,
  breaklines=true,
}
\lstdefinestyle{mintedclone2}{
  language=C++,
  basicstyle=\ttfamily\scriptsize,
  numbers=left,
  numberstyle=\scriptsize\color{gray},
  frame=lines,
  rulecolor=\color{gray!60},
  keywordstyle=\color{blue!70!black},
  commentstyle=\color{gray!70},
  stringstyle=\color{orange!80!black},
  showstringspaces=false,
  tabsize=2,
  breaklines=true,
}

\definecolor{background}{RGB}{250,250,250}
\definecolor{string}{RGB}{206,123,0}
\definecolor{number}{RGB}{0,102,204}
\definecolor{key}{RGB}{0,128,0}
\definecolor{boolean}{RGB}{163,21,21}

\lstdefinelanguage{JSON}{
    basicstyle=\ttfamily\small,
    numbers=left,
    numberstyle=\tiny\color{gray},
    stepnumber=1,
    numbersep=5pt,
    showstringspaces=false,
    breaklines=true,
    breakatwhitespace=false,
    backgroundcolor=\color{background},
    literate=
     *{"}{{\textcolor{string}{"}}}1
      {0}{{\textcolor{number}{0}}}1
      {1}{{\textcolor{number}{1}}}1
      {2}{{\textcolor{number}{2}}}1
      {3}{{\textcolor{number}{3}}}1
      {4}{{\textcolor{number}{4}}}1
      {5}{{\textcolor{number}{5}}}1
      {6}{{\textcolor{number}{6}}}1
      {7}{{\textcolor{number}{7}}}1
      {8}{{\textcolor{number}{8}}}1
      {9}{{\textcolor{number}{9}}}1
      {true}{{\textcolor{boolean}{true}}}4
      {false}{{\textcolor{boolean}{false}}}5
      {null}{{\textcolor{boolean}{null}}}4
      {:}{{\textcolor{key}{:}}}1
      {,}{{\textcolor{key}{,}}}1
}

\definecolor{excellent}{HTML}{DD6893}     
\definecolor{verygood}{HTML}{E48EAC}      
\definecolor{good}{HTML}{F0B8C8}          
\definecolor{neutral}{RGB}{189,189,189}   
\definecolor{tableback}{HTML}{FFF8E7}     

\definecolor{accent}{HTML}{DD6893}        
\definecolor{accentLight}{HTML}{E48EAC}   
\definecolor{accentLighter}{HTML}{F3C2D4} 
\definecolor{neutral}{RGB}{189,189,189}   
\definecolor{tableback}{HTML}{FFF9F5}     
\definecolor{headerback}{HTML}{FBE7EE}    

\tcbset{
  tableframe/.style={
    colback=white,
    colframe=accent!60!black,
    boxrule=0.6pt,
    arc=3mm,
    left=3mm,
    right=3mm,
    top=2mm,
    bottom=2mm,
    enhanced,
  }
}

\newcolumntype{C}[1]{>{\centering\arraybackslash}p{#1}}

\title{\texttt{irace-evo}: Automatic Algorithm Configuration Extended With LLM-Based Code Evolution}
\author{
  Camilo Chacón Sartori \orcidlink{0000-0002-8543-9893}\thanks{Corresponding author.}\\
  Catalan Institute of Nanoscience and Nanotechnology (ICN2),\\
  CSIC and BIST, Campus UAB, Bellaterra, 08193 Barcelona, Spain\\ 
  \texttt{cchacon@icn2.cat} \\
  Artificial Intelligence Research Institute (IIIA-CSIC)\\
  Bellaterra, Spain\\ 
  \texttt{cchacon@iiia.csic.es} \\
  \and
  Christian Blum \orcidlink{0000-0002-1736-3559}\\
  Artificial Intelligence Research Institute (IIIA-CSIC)\\
  Bellaterra, Spain\\
  \texttt{christian.blum@iiia.csic.es}
}

\begin{document}
\maketitle      

\begin{figure}[h]
\centering
  \includegraphics[width=0.7\linewidth]{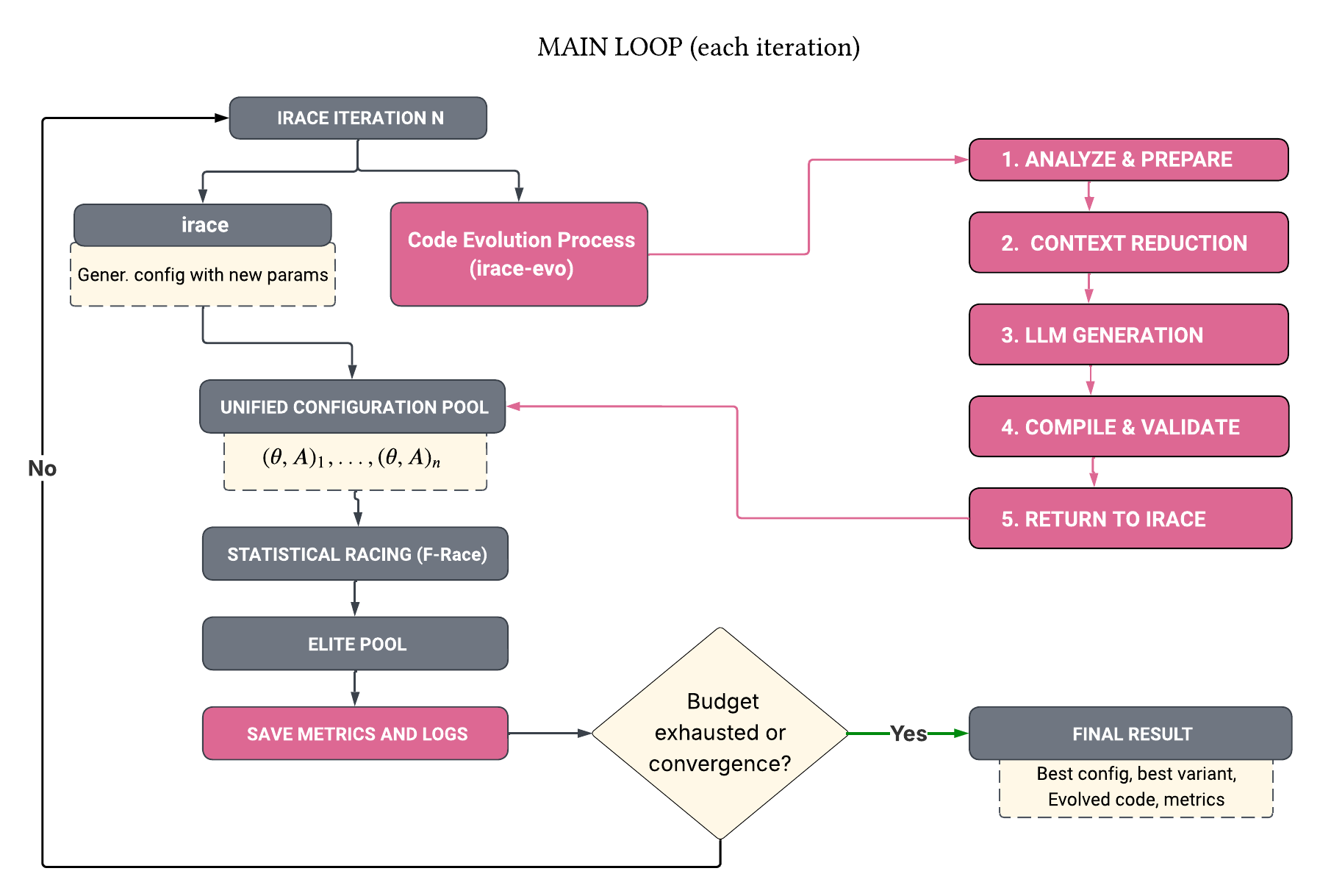}
  \caption{A flowchart showing the integration of \texttt{irace-evo} into \texttt{irace}.}
  \label{fig:example}
\end{figure}

\begin{abstract}
Automatic algorithm configuration tools such as \texttt{irace} efficiently tune parameter values but leave algorithmic code unchanged. This paper introduces a first version of \texttt{irace-evo}, an extension of \texttt{irace} that integrates code evolution through large language models (LLMs) to jointly explore parameter and code spaces. The proposed framework enables multi-language support (e.g., C++, Python), reduces token consumption via progressive context management, and employs the Always-From-Original principle to ensure robust and controlled code evolution. We evaluate \texttt{irace-evo} on the Construct, Merge, Solve \& Adapt (CMSA) metaheuristic for the Variable-Sized Bin Packing Problem (VSBPP). Experimental results show that \texttt{irace-evo} can discover new algorithm variants that outperform the state-of-the-art CMSA implementation while maintaining low computational and monetary costs. Notably, \texttt{irace-evo} generates competitive algorithmic improvements using lightweight models (e.g., Claude Haiku 3.5) with a total usage cost under \euro 2. These results demonstrate that coupling automatic configuration with LLM-driven code evolution provides a powerful, cost-efficient avenue for advancing heuristic design and metaheuristic optimization.
\end{abstract}
%
%
%


\section{Introduction}

The configuration of a stochastic optimization algorithm can significantly impact its performance. Achieving satisfactory results, therefore, requires selecting appropriate parameter values, which may vary across different instances. This is a process known as parameter tuning, an active field of research. Algorithm parameters can differ in data type---such as continuous, categorical, or Boolean---and in the range or domain of values that influence the algorithm’s behavior. For example, a genetic algorithm might have parameters such as population size, mutation rate, and number of elite solutions, whereas an ant colony optimization algorithm, among others, has parameters like the number of ants and the pheromone decay rate. Poorly chosen parameter values can severely degrade performance. In essence, parameter tuning represents an optimization problem, in which finding the optimal configuration of $n$ parameters remains a challenging task.

The related literature offers several tools for automatic parameter tuning. Apart from \texttt{irace}~\cite{LOPEZIBANEZ201643}, which will be described in Section~\ref{sec:irace-intro}, often used tools are \texttt{SMAC}~\cite{10.1007/978-3-642-25566-3_40} and \texttt{ParamILS}~\cite{10.5555/1734953.1734959}, which use model-based and local-search strategies, respectively. Another tool called \texttt{Hyperband}~\cite{DBLP:journals/jmlr/LiJDRT17} uses early stopping to quickly allocate resources, while \texttt{BOHB}~\cite{pmlr-v80-falkner18a} combines Bayesian optimization with \texttt{Hyperband}’s resource allocation to balance exploration and exploitation. Both are effective but may require more iterations to converge than \texttt{irace}.

\paragraph{Our contribution.} The above-mentioned tools are capable of optimizing the parameter values of a given algorithm, but do not modify the code itself. In contrast, in this work, we present \texttt{irace-evo} (see Figure~\ref{fig:irace-evo}), an extension of \texttt{irace} to explore both the parameter value space and the code space simultaneously, enabling co-evolution of the algorithm and its parameter values. Code changes are performed in \texttt{irace-evo} with the help of large language models (LLMs). We show that \texttt{irace-evo}, even with few iterations, can effectively guide this dual optimization process. An example application to a CMSA algorithm for the so-called Variable-Sized Bin Packing Problem (VSBPP) (published in~\cite{Akbay2024}) shows that \texttt{irace-evo} is able to improve current state-of-the-art algorithms even further.

\subsection{irace: Iterated Racing for Automatic Algorithm Configuration}
\label{sec:irace-intro}

In metaheuristics, one of the most widely used tools for automatic parameter configuration is \texttt{irace}~\cite{LOPEZIBANEZ201643}. Based on the F-Race procedure~\cite{10.5555/2955491.2955494}, \texttt{irace} employs an \textit{iterated racing} strategy that systematically evaluates candidate parameter configurations across multiple problem instances. It begins with a large pool of configurations and progressively discards inferior ones as soon as statistical evidence---via the Friedman rank test---indicates significant performance differences. A minimum number of evaluations is required before eliminations, reducing computational cost by avoiding poor candidates early.

The process is iterative: initial configurations are sampled randomly from the parameter space, evaluated on instances, and the best (elite) ones are used to model promising regions for the next sampling round. Enhancements such as restarts, truncated sampling distributions, and elitist selection improve robustness and efficiency. The procedure continues until the computational budget is exhausted, returning the best elite configurations.

The performance of \texttt{irace} depends strongly on the quality and diversity of the training instances. If they are too simple or homogeneous, \texttt{irace} may fail to detect meaningful statistical differences, producing configurations that generalize poorly to more complex or varied problem instances.

\subsection{Related Work}

LLMs have revitalized optimization in many aspects including self-improving code, evolutionary programming, program synthesis, algorithm discovery, and automated tuning. Leveraging their transformer architecture, LLMs can generate problem-specific heuristics, adapt algorithmic code, and evolve solutions iteratively, providing new impetus to optimization algorithms while benefiting from their capability to produce coherent code.

Recent work has explored LLM-driven algorithm design. A prominent example is FunSearch~\cite{Romera-Paredes2024} combining LLMs with evolutionary search for solving combinatorial optimization problems. Furthermore, Evolution of Heuristics (EoH)~\cite{liu2024evolutionheuristicsefficientautomatic} is a system in which algorithms evolve through LLM-guided mutations. LLaMEA~\cite{10752628} generates metaheuristics from scratch in Python by means of LLM-based evolution and self-debugging, and LLaMEA-HPO~\cite{10.1145/3731567} separates code generation from parameter tuning via SMAC~\cite{lindauer2022smac3versatilebayesianoptimization}, achieving up to 20× fewer LLM queries while maintaining competitive performance on benchmarks like Online Bin Packing, Black-Box Optimization, and the TSP.

Despite their promise, these frameworks share limitations: they are restricted to Python, incur high token costs by sending complete algorithm codes to the LLM at each iteration, and lack systematic mechanisms for detecting and recovering from evolutionary computation issues such as premature convergence.


\section{\texttt{irace-evo}: Adding Code Evolution to \texttt{irace}}

In this section, we introduce \texttt{irace-evo}, an extension of \texttt{irace} for addressing some of the challenges of existing approaches outlined above. We do so with a multi-language architecture, progressive context management, and automated performance analysis. Fully compatible with \texttt{irace}~\cite{LOPEZIBANEZ201643}, \texttt{irace-evo} extends it to support LLM-driven code evolution, jointly exploring parameter configuration and algorithm code spaces to discover high-performing parameter settings while improving user-provided code, bridging automated configuration and evolutionary code generation.

\subsection{Design Principles}

Building on insights from recent LLM-driven frameworks, \texttt{irace-evo} addresses four main challenges:

\begin{enumerate}
\item \textbf{Multi-Language Support.} Unlike existing Python-based frameworks (EoH, LLaMEA, LLaMEA-HPO), our plugin-based system supports both C++ and Python, and can be extended to other programing languages.

\item \textbf{Reducing Tokens.} Beyond LLaMEA-HPO’s HPO separation, we use progressive context management to reduce token usage by 60–80\%.

\item \textbf{Flexible Configuration.} The use of \texttt{irace} instead of \texttt{SMAC} provides conditional parameters and multi-language support without Python-specific constraints.

\item \textbf{Efficient LLM Usage:} Instead of relying on large, costly LLMs, \texttt{irace-evo} uses efficient, low-cost models to surpass existing heuristics.
\end{enumerate}

\noindent These principles are implemented in the architecture described next.

\subsection{Architecture Overview}

The operating mode of \texttt{irace-evo} was designed to remain fully compatible with \texttt{irace}~\cite{LOPEZIBANEZ201643}. In fact, \texttt{irace-evo} (instead of standard \texttt{irace}), is evoked by a simple addition to the configuration file of \texttt{irace}:

\begin{lstlisting}[style=mintedclone]
## Code Evolution Configuration
codeEvolution = "TRUE"
codeEvolutionConfig = ".~/code-evolution.json"
codeEvolutionVariants = 5
\end{lstlisting}

In particular, setting parameter \texttt{codeEvolution} to \texttt{TRUE} evokes the functionalities of \texttt{irace-evo}. Furthermore, the path and file name stored in parameter \texttt{codeEvolutionConfig} indicate to \texttt{irace-evo} where to find the complete configuration, including the source code of the algorithm $A$ to be tuned, the description of the tackled optimization problem, and the indication of the part of the code that the LLM is allowed to change.\footnote{Currently, \texttt{irace-evo} is allowed to change exactly one function of the provided algorithm code. In our experience, the more well-defined this part is, the higher is the probablity that \texttt{irace-evo} can find an improved algorithm code.} Finally, parameter \texttt{codeEvolutionVariants} indicates the number of code variants to be produced and tested by \texttt{irace-evo} at each iteration. \\

However, before describing the extensions coming with \texttt{irace-evo}, let us first recall the standard \texttt{irace} funcionality. Formally, \texttt{irace} addresses the problem of configuring the parameters of a given optimization algorithm $A$ by means of a search for the parameter configuration $\theta^* \in \Theta$ that minimizes the expected cost function $F()$ over a distribution of problem instances $\mathcal{I}$. Since $F(\theta)$ for some $\theta \in \Theta$ cannot be evaluated analytically, it is estimated empirically by repeatedly applying algorithm $A$ with parameter configuration $\theta$ to instances sampled from an instance distribution $\mathcal{I}$ (generally given in terms of a set of tuning instances) under a limited computational budget of $B$ algorithm runs. In fact, \texttt{irace} approximates this optimization process through an iterative \emph{racing} mechanism: at each iteration $t$, a set of candidate algorithm configurations $\texttt{Conf} = \{\theta_1, \ldots, \theta_{N_t}\}$ is sampled from a probabilistic model $P$ over $\Theta$, evaluated on a set of instances, and statistically compared using non-parametric tests (e.g., Friedman or paired $t$-test). Inferior candidates are discarded, while surviving \emph{elite configurations} $\texttt{Conf}_{\mathrm{elite}}$ are used to update $P$ towards promising regions:
\begin{equation}
  P = \text{UpdateModel}(P, \texttt{Conf}_{\mathrm{elite}})
\end{equation}
This process continues until the budget $B$ is exhausted, yielding the final best configuration $\hat{\theta}$. \\

In the following, the algorithm code provided by the user will be called $A^0$, referring to the original variant (or variant zero) of the algorithm to be optimized. Moreover, $\mathcal{A}$ denotes the space of all possible algorithm variants that can be generated by changing, in some way, the code of the function of $A^0$ that is subject to change, as specified in the JSON file indicated by parameter \texttt{codeEvolutionConfig} of the \texttt{irace} configuration file.

\LinesNumbered
\begin{algorithm}[!t]
\scriptsize
\SetAlgoNlRelativeSize{-1}
\SetAlgoNlRelativeSize{0}
\caption{Pseudo-code of \texttt{irace-evo}}\label{algo:irace-evo}
\KwIn{Initial parameter space $\Theta$, budget $B$}
\KwOut{Final best configuration $(\hat{\theta}, \hat{A})$}
$\texttt{Conf}_{\mathrm{elite}} \gets \emptyset$  \;
$P \gets$ Uniform random distribution over $\Theta$ \;
\While{budget $B$ not exhausted and not converged}{
    \tcc{Generate new configurations}
    $\mathcal{A}_{\mathrm{survived}} \gets \{ A \mid \exists (\theta, A) \in \texttt{Conf}_{\mathrm{elite}}\}$ \;
    \colorbox{evo}{$\mathcal{A}_{\mathrm{evo}} \gets$ \texttt{GenerateCodeVariants}(} \;
    \colorbox{evo}{$\mathcal{A}_{\mathrm{valid}} \gets$ \texttt{Validate}($\mathcal{A}_{\mathrm{evo}}$)} \;
    \colorbox{evo}{$\texttt{Conf}_{\mathrm{new}} \gets$ \texttt{GenerateNewConfigurations}($\mathcal{A}_{\mathrm{survived}}, A^0, \mathcal{A}_{\mathrm{valid}}, P$)} \;
    $\texttt{Conf}_{\mathrm{pool}} \gets \texttt{Conf}_{\mathrm{elite}} \cup \texttt{Conf}_{\mathrm{new}}$ \;
    
    \tcc{Evaluate and select best candidates using F-Race}
    \ForEach{$(\theta, A) \in \texttt{Conf}_{\mathrm{pool}}$}{
        Evaluate $(\theta, A)$ \;
    }
    $\texttt{Conf}_{\mathrm{elite}} \gets$ F-Race-Select($\texttt{Conf}_{\mathrm{pool}}$) \;
    \tcc{Update the probablistic model over $\Theta$}
    $P \gets \text{UpdateModel}(P, \texttt{Conf}_{\mathrm{elite}})$ \;
}

\tcc{Return the final best configuration and code variant}
$(\hat{\theta}, \hat{A}) \gets$ \texttt{SelectBest}($\texttt{Conf}_{\mathrm{elite}}$) \;
\Return $(\hat{\theta}, \hat{A}) \gets$
\end{algorithm}

\begin{figure}[!t]
\centering
  \includegraphics[width=0.7\linewidth]{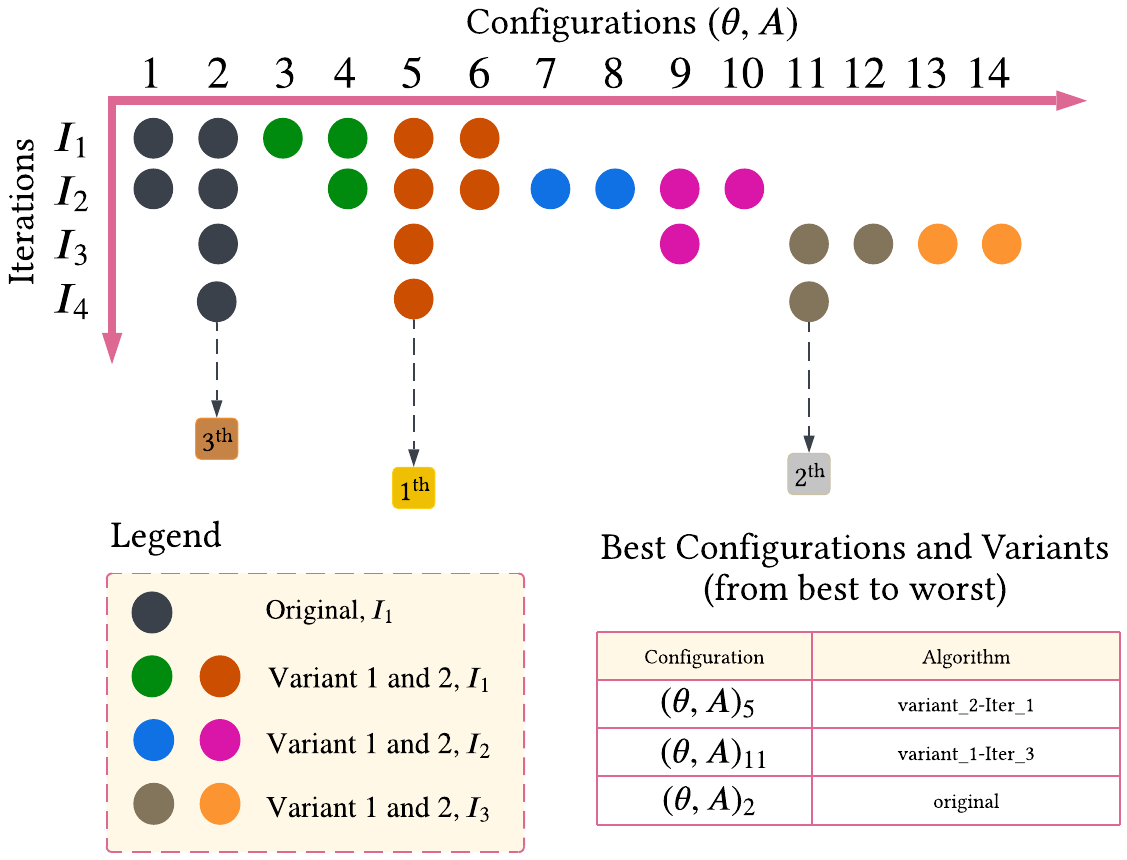}
    \caption{Example of an \texttt{irace-evo} run with \texttt{codeEvolutionVariants=2}.}
  \label{fig:irace-evo}
\end{figure}

The extension from \texttt{irace} to \texttt{irace-evo} consists in extending the definition of a configuration from the parameter value space $\Theta$ to combinations between parameter configuration and an algorithm code variant; that is, configurations in \texttt{irace-evo} are $(\theta, A)$ where $\theta \in \Theta$ and $A \in \mathcal{A}$. Algorithm~\ref{algo:irace-evo} illustrates the pseudo-code of \texttt{irace-evo}, highlighting in pink the operations specific to code evolution. The workflow is as follows: first, in lines~1 and~2, the pool of elite configurations is initialized to the empty set, and $P$ is initialized to the uniform random distribution over $\Theta$. Then, at the beginning of each iteration (see line~4), $\mathcal{A}_{\mathrm{survived}}$ is determined as the set of all algorithm variants that appear in at least one configuration of the elite pool $\texttt{Conf}_{\mathrm{elite}}$. Next, in line~5, new algorithm variants are produced by the utilized LLM in function \texttt{GenerateCodeVariants}(), as explained below, and they are stored in $\mathcal{A}_{\mathrm{evo}}$.\footnote{Remember that the number of newly generated algorithm variants is determined by parameter \texttt{codeEvolutionVariants} in the configuration file of \texttt{irace}.} After that, all new algorithm code variants are validated (see line~6) and only those that compile without errors are kept in set $\mathcal{A}_{\mathrm{valid}}$. New configurations are then generated in function \texttt{GenerateNewConfigurations}(.) (see line~7) on the basis of sets $\mathcal{A}_{\mathrm{survived}}$ and $\mathcal{A}_{\mathrm{valid}}$, as well as $A^0$ and distribution $P$, as follows. First, an algorithm variant $A$ is chosen from $\mathcal{A}_{\mathrm{survived}} \cup \mathcal{A}_{\mathrm{valid}} \cup \{A^0\}$ and, second, a parameter configuration $\theta$ is sampled from $P$. The resulting configuration $(\theta, A)$ is returned and added to $\texttt{Conf}_{\mathrm{new}}$. This is done until $\texttt{Conf}_{\mathrm{pool}} \cup \texttt{Conf}_{\mathrm{new}}$ has the desired size (which is internally calculated by \texttt{irace}), and $\texttt{Conf}_{\mathrm{new}}$ is added to $\texttt{Conf}_{\mathrm{pool}}$. Finally, all confiugrations from $\texttt{Conf}_{\mathrm{pool}}$ are evaluated and a new elite set of configurations $\texttt{Conf}_{\mathrm{elite}}$ is selected from $\texttt{Conf}_{\mathrm{pool}}$ in the standard way of \texttt{irace}; see lines~9-12. A graphical illustration of this process is shown in Figure~\ref{fig:irace-evo}.

\subsection{LLM Integration}

The function \texttt{GenerateCodeVariants}() (see Algorithm~\ref{algo:irace-evo}, line~5) is central to \texttt{irace-evo}. Unlike other frameworks that evolve code cumulatively---building each new variant from previously generated versions---\texttt{irace-evo} takes a different approach. We assume that the original algorithm variant $A^0$ is already of high quality (e.g.~state-of-the-art). Therefore, \texttt{irace-evo} adopts the Always-From-Original (AFO) principle: each new variant $\mathcal{A}_{\mathrm{evo}}$ is generated directly from $A^0$, rather than from earlier evolved versions.

This strategy preserves the baseline performance while allowing controlled exploration of alternative improvements. By always referencing $A^0$, the system avoids the accumulation of errors and unintended structural changes, ensuring that each generated variant is both safe and potentially beneficial. Formally,
\[
\mathcal{A}_{\text{evo}} = \{\, \texttt{LLM}(\texttt{PROMPT}_i) \mid i = 1, \dots, n \,\} \enspace,
\]
where $n$ refers to the user-provided number stored in parameter \texttt{codeEvolution- Variants} in the configuration file of \texttt{irace}, and where each prompt $\texttt{PROMPT}_i$ is dynamically constructed as follows:
\[
\texttt{PROMPT}_i = f\big(FC, \; \mathcal{E}_i, \; R_i, \; \mathcal{J}, \; \Phi \big) \enspace,
\]
with:
\begin{itemize}
    \item $FC$: the code of the function of the original code $A^0$ which the LLM is allowed to change;
    \item $\mathcal{E}_i$: a prompt segment controlling how aggressively the LLM modifies  $A^0$; it has two modes $\{\texttt{std}, \texttt{aggressive}\}$, and a parameter $ef \geq 1$ is used that specifies the iteration at which the behavior change from \texttt{std} to \texttt{agressive} occurs;
    \item $R_i$: the code context window, representing either the entire code of $A^0$ or a subset thereof, progressively reduced to optimize token usage;
   \item $\mathcal{J}$: a structured JSON object containing the \textit{problem context}---including the problem name, description, algorithmic approach, optimization objective, domain knowledge, and key challenges. This is the file referenced in the \texttt{codeEvolutionConfig} field of the \texttt{irace} configuration.
    \item $\Phi$: the LLM hyperparameters (temperature and top-$p$).
\end{itemize}
The roles of $\mathcal{E}_i$, $R_i$, and $\mathcal{J}$ are further detailed in Section~\ref{subsec:dynamic-prompts}. We finish this subsection by outlining the three key advantages of the adopted design: (1) it \textit{avoids running into local optima} and, therefore, does not suffer from an accumulation of suboptimal variants; (2) it \textit{ensures broad exploration} each iteration independently starts from the original algorithm variant $A^0$; (3) it is likely to be \textit{robust} because performance does not rely on early generations. Consequently, \texttt{irace-evo} performs a \textit{short code evolution}, emphasizing subtle yet impactful modifications to the original variant $A^0$---a phenomenon empirically validated in Section~\ref{sec:experiments}.

\subsection{Dynamic Prompts}\label{subsec:dynamic-prompts}

\subsubsection{Evolution Focus ($\mathcal{E}_i$)}
This component controls the intensity of LLM modifications, operating in two modes: \texttt{std}, which refers to general algorithm improvement by moderate changes preserving the overall logic, and \texttt{aggressive}, which aims at a more aggressive exploration, that is, broader structural changes. Each run of \texttt{irace-evo} starts in the first mode and switches to the second after a predefined iteration threshold $ef$ (e.g., $ef=3$).

\subsubsection{Code Context Window ($R_i$)}
To reduce token usage, \texttt{irace-evo} applies progressive context management: full-context initialization at iteration $i=1$ (entire source file of $A^0$) and selective fragments for $i>1$ (e.g., global variables, relevant \texttt{structs}, imports). This is done because excessive or irrelevant context may degrade performance, enabling the study of the trade-off between contextual richness and token efficiency.

\subsubsection{Problem and Algorithm Context ($\mathcal{J}$)}
A JSON configuration file specifies detailed information about the optimization problem and the provided algorithm $A^0$, influencing LLM code generation. Let $\mathcal{J}$ denote this context and $Q(A)$ the variant quality. In general, richer, relevant context improves variant quality ($Q(A) \propto |\mathcal{J}|$), though excessive or poorly curated information may introduce noise. This allows \texttt{irace-evo} to balance context relevance and output quality.

\subsection{Implementation Details}

\textbf{Integrating Code Evolution into \texttt{irace}.} \texttt{irace-evo} integrates Python into R via the \texttt{reticulate} package, enabling LLM access without modifying core \texttt{irace} functionality. Only three files are extended: \texttt{R/irace-options.R} (load configurations), \texttt{R/irace.R} (adapt execution flow), and \texttt{R/scenario.R} (add validation routines), preserving backwards compatibility while enabling code evolution. \\

\textbf{Plugin-Based Language Integration.} A plugin-based architecture allows \texttt{irace-evo} to support multiple languages. Algorithm execution relies on two operations: \emph{search} and \emph{replacement}. A partial parser identifies function signatures and bodies (search) and replaces the function it is allowed to change with LLM-generated code (replacement). For instance, the C++ parser that we developed for our example application described in Section~\ref{sec:example-application} detects function definitions, sufficient for replacement. A user who wants to use \texttt{irace-evo} for an algorithm coded in a different language needs to develop a corresponding parser. \\

\textbf{Verification and Validation.} \emph{Verification} ensures executable correctness: compilation is retried up to $k$ times, and runtime errors are penalized. \emph{Validation} ensures the evolved code produces meaningful results. Since only specific code regions (currently: one specific function) are modified, validation quality depends on how well the target function is structured; functions combining multiple algorithmic components or constraints increase error risk and may reduce the performance of new variants of $A^0$.

\section{Case Study: CMSA for the VSBPP Problem}
\label{sec:example-application}

The following case study evaluates \texttt{irace-evo} by applying it to a Construct, Merge, Solve \& Adapt (CMSA) metaheuristic for the Variable-Sized Bin Packing Problem (VSBPP)~\cite{Akbay2024}. Note that this algorithm is currently the state of the art for this problem.

\subsection{Problem Definition}

The VSBPP is a combinatorial optimization problem formally defined as follows. Given is a set $S = \{1, \dots, n\}$ of $n$ items, where each item $i \in S$ has a positive weight $w_i > 0$. Furthermore, a set $B = \{1, \dots, m\}$ of $m$ available bin types is given. Each bin type $k \in B$ is characterized by a capacity $W_k > 0$ and an associated cost $C_k$. The objective is to pack all $n$ items into a selection of bins such that the sum of the weights of the items in each bin does not exceed the bin type's capacity. The goal is to minimize the total cost of all bins used. There is no restriction on the number of times a bin type may be used.

\subsection{The Base Algorithm CMSA}

CMSA is an iterative hybrid metaheuristic framework~\cite{10.5555/2873828.2874036}. At each iteration, it solves a subinstance of the original problem, whose search space is a subset of the full problem's search space. The algorithm repeatedly executes four main steps:
\begin{itemize}
    \item \textbf{Construct:} A set of valid solutions to the problem is generated, often using a probabilistic greedy heuristic.
    \item \textbf{Merge:} The solution components (e.g., valid bin sets) found in these solutions are collected and added to the incumbent subinstance, hereby expanding it.
    \item \textbf{Solve:} The current subinstance is solved using an exact or heuristic solver (such as an ILP solver) to find a high-quality, feasible solution using only the components present in the subinstance.
    \item \textbf{Adapt:} The subinstance is reduced by removing components, typically using an aging mechanism. Components that are not part of the solutions found in the Solve-step for several consecutive iterations are removed.
\end{itemize}

The CMSA algorithm described in~\cite{Akbay2024} for the VSBPP problem is implemented in C++. A relevant consideration is that, for performance reasons, many optimization algorithms are implemented in C++. Unlike interpreted programming languages such as Python, C++ introduces no runtime overhead from dynamic typing or interpreter design, making it particularly suitable for high-efficiency optimization implementations. \\

\subsubsection{From Heuristic to Short Evolution}\label{subsubsec:original-heuristic}

The given C++ code of CMSA makes heuristic decisions at several steps. In the Construct-Step, for example, solutions are generated by placing items one-by-one, either into an already opened bin or after opening a new bin. The choice of the bin in which to place the current item is hereby based on a straightforward cost-to-load ratio as implemented in function \texttt{evaluate\_placement\_quality}(); see below. Although being less sophisticated than alternative criteria, with this implementation, the algorithm achieves high-quality solutions through repeated iterations, making it hard to replace---since even a “better” heuristic criterion might underperform under a tight computational time limit. Thus, any alternative criterion (or heuristic) to be evolved for making this placement decision must be an improved heuristic while preserving computational efficiency.

\begin{lstlisting}[style=mintedclone2]
// Independent function for evaluating placement quality
// Receives detailed information to implement intelligent evaluation metrics
double evaluate_placement_quality(int new_bin_type, 
                                 int new_load,
                                 const vector<int>& bin_costs) {
    // Current heuristic: cost/load ratio
    return bin_costs[new_bin_type] / double(new_load);
}
\end{lstlisting}
\section{Results and Analysis}\label{sec:experiments}

This section presents the results of applying \texttt{irace-evo} to the C++ implementation of CMSA for the VSBPP. In the following, we describe the experimental setup and analyze the outcome of the conducted experiments.

\subsection{Experimental Setup}

The base algorithm (CMSA) is implemented in C++ and was tested on a cluster equipped with Intel Xeon 5670 CPUs (12 cores at 2.933~GHz) and at least 32~GB of RAM. Subinstances are solved in the Solve-step of CMSA using CPLEX~22.1 configured to run on a single thread.

\begin{itemize}
\item \textbf{Benchmark.} We used the same set of benchmark instances as in~\cite{Akbay2024}. Instances in this set comprise seven bin types with capacities $W = \{70, 100, 130,$ $160, 190, 220, 250\}$ and item weights uniformly sampled from $[1, 250]$. Three instance classes are considered: (B1) linear bin costs $C_i = W_i$, (B2) concave bin costs $C_i = \lceil10\sqrt{W_i}\rceil$, and (B3) convex bin costs $C_i = \lceil0.1 W_i^{3/2}\rceil$, for $i = 1, \dots, 7$. Each class contains ten instances for $n \in {100, 200, 500, 1000, 2000}$, totaling 150 problem instances.\footnote{Optimal solutions for this benchmark are not known.}

\item \textbf{irace-evo configuration.} Due to the LLM-incurred costs, we restrict the experimental budget $B$ to an unusually low number of (\texttt{maxExperiments} = 300) to test whether \texttt{irace-evo} can discover improved heuristics within a low number of iterations. Moreover, each algorithm run was given a time limit of 150 CPU seconds.

\item \textbf{Code evolution.} In each \texttt{irace-evo} iteration, five heuristic variants are generated from the original heuristic using dynamically crafted prompts~(see Subsection~\ref{subsec:dynamic-prompts}). We use \texttt{Claude~Haiku~3.5} (Anthropic) for its balance between efficiency and quality, operating with $T = 1.0$ and $\text{top}\_p = 0.9$ (LLM hyperparameters $\Phi$)  to yield diverse yet coherent algorithm variants under a limited computational budget.

\item \textbf{Comparison.} After ten independent runs of \texttt{irace-evo}, we obtain ten distinct algorithm variants, that is, ten variants of the original algorithm variant $A^0$, all differing in the implementation of function \texttt{evaluate\_placement \_quality}(.). Each algorithm variant was applied (with a time limit of 150 CPU seconds) to each instance of classes B1, B2, and B3. The resulting performance is compared against the results of the original algorithm variant from~\cite{Akbay2024}. This is discussed in detail in the following subsection.
\end{itemize}

\subsection{Analysis}

As the algorithm variants differ in the placement heuristic, we henceforth simply denote them by $H_1,\dots,H_{10}$. Table~\ref{tab:winrate_comparison} presents a statistical comparison of $H_1,\dots,H_{10}$ against the original algorithm variant. To ensure a fair comparison\footnote{Remember that the computational budget $B$ of \texttt{irace-evo} was chosen quite low due to LLM-incurred costs.} with the original algorithm variant, each of the ten algorithm variants underwent a subsequent parameter tuning by \texttt{irace} (just like the original algorithm variant from~\cite{Akbay2024}). In fact, in Table~\ref{tab:winrate_comparison}, \textit{Std} refers to the algorithm variants using the parameter setting as determined by \texttt{irace-evo}, while \textit{Tun} refers to the subsequently tuned version. The observed behavior across datasets is as follows:

\begin{itemize}
\item \textbf{B3.} \texttt{irace-evo} algorithm variants generally outperform the original algorithm on larger instances (especially $n = 1000$ and $2000$), with slightly better results when tuned. However, $H_4$, $H_8$, $H_9$, and $H_{10}$ perform worse than the other ones, even though they are improved by tuning. For smaller instances ($n = 100$ and $200$), performance is similar to the original algorithm variant, indicating that larger instances are needed to observe meaningful gains.

\item \textbf{B2.} The advantage of the algorithm variants produced by \texttt{irace-evo} is lower than in the case of B3 but still statistically significant. $H_5$ achieves the best results with and without tuning, while $H_4$ performs better than for B3-instances, suggesting that the type of cost function used makes a difference.

\item \textbf{B1.} The statistical advantage over the original algorithm variant is further reduced, with only $H_7$ showing a clear improvement.

\end{itemize}

Overall, these results show that, despite differences in instance characteristics among B1, B2, and B3, \texttt{irace-evo} consistently produces algorithm variants that outperform the original algorithm---subsequently fine-tuned or not---with the largest gains seen in B3. \\

Next, in Table~\ref{tab:improvements}, we analyze which algorithm variants are able to produce new best solutions in comparison to the original algorithm from~\cite{Akbay2024}. The results reveal a consistent pattern: Most new best solutions are found for B3 instances, and the subsequently fine-tuned algorithm variants generally produce, on average, slightly better solutions. Yet, the untuned versions also perform remarkably well, showing that \texttt{irace-evo} can generate competitive heuristics even without a full-fledged parameter tuning (remember that the computational budget of the \texttt{irace-evo} runs was chosen very low). Moreover, the consistency across the ten independent \texttt{irace-evo} runs demonstrates that \texttt{irace-evo} reliably produces high-quality algorithm variants, with several outstanding ones (e.g., $H_5$ and $H_7$). 

\begin{table}[!t]
  \centering
  \renewcommand{\arraystretch}{1}
  \setlength{\tabcolsep}{3pt}
  \small

  \caption{Win rates in comparison to the original algorithm variant. The win rate (in \%) indicates the percentage of instances for which an algorithm variant outperforms the original algorithm. Statistical significance is indicated as follows: $^{***}$ indicate a $p$-value of $<0.001$, $^{**}$ one of $<0.01$, and $^{*}$ one of $<0.05$. 
Color coding:
\fcolorbox{black}{excellent}{\raisebox{0.7ex}{\phantom{.}}} ~Excellent (win rate $>$50\%, highly significant),
\fcolorbox{black}{verygood}{\raisebox{0.7ex}{\phantom{.}}}~Very good (20--50\%, significant),
\fcolorbox{black}{good}{\raisebox{0.7ex}{\phantom{.}}}~Good (significant improvement),
\fcolorbox{black}{neutral}{\raisebox{0.7ex}{\phantom{.}}}~Neutral (0\%, poor performance),
\fcolorbox{black}{tableback}{\raisebox{0.7ex}{\phantom{.}}}~Non-significant improvement.}
  \label{tab:winrate_comparison}

  \vspace{0.3em}
  \begin{adjustbox}{max width=\textwidth}
  \begin{tabular}{@{}ll|>{\columncolor{tableback}}c>{\columncolor{tableback}}c>{\columncolor{tableback}}c>{\columncolor{tableback}}c>{\columncolor{tableback}}c|>{\columncolor{tableback}}c>{\columncolor{tableback}}c>{\columncolor{tableback}}c>{\columncolor{tableback}}c>{\columncolor{tableback}}c|>{\columncolor{tableback}}c>{\columncolor{tableback}}c>{\columncolor{tableback}}c>{\columncolor{tableback}}c>{\columncolor{tableback}}c@{}}
  \toprule[1.5pt]
  \multicolumn{2}{c|}{\textbf{Algorithm}} & \multicolumn{5}{c|}{\textbf{B1}} & \multicolumn{5}{c|}{\textbf{B2}} & \multicolumn{5}{c}{\textbf{B3}} \\
  \cmidrule(lr){3-7} \cmidrule(lr){8-12} \cmidrule(l){13-17}
  \multicolumn{2}{c|}{\textbf{Variant}} & \textbf{100} & \textbf{200} & \textbf{500} & \textbf{1000} & \textbf{2000} & \textbf{100} & \textbf{200} & \textbf{500} & \textbf{1000} & \textbf{2000} & \textbf{100} & \textbf{200} & \textbf{500} & \textbf{1000} & \textbf{2000} \\
  \midrule[1.2pt]

  \multirow{2}{*}{\textbf{H1}} & \textit{Std} & \cellcolor{neutral}0*** & \cellcolor{neutral}0*** & \cellcolor{neutral}0*** & \cellcolor{good}14** & 31 & 0 & \cellcolor{good}9** & \cellcolor{good}8* & 28 & \cellcolor{good}9*** & 0 & 0 & \cellcolor{good}10** & \cellcolor{excellent}63*** & \cellcolor{excellent}88*** \\
  & \textit{Tun} & \cellcolor{neutral}0*** & \cellcolor{neutral}0*** & \cellcolor{neutral}0*** & \cellcolor{good}18* & 32 & 0 & \cellcolor{good}6* & 6 & \cellcolor{good}35* & \cellcolor{good}17* & 0 & 0 & \cellcolor{good}10** & \cellcolor{excellent}63*** & \cellcolor{excellent}90*** \\
  \midrule

  \multirow{2}{*}{\textbf{H2}} & \textit{Std} & 0 & 0 & 1 & \cellcolor{good}17* & 38 & 0 & \cellcolor{good}9** & 8 & 23 & \cellcolor{verygood}11*** & 0 & 0 & 1 & \cellcolor{excellent}53*** & \cellcolor{excellent}92*** \\
  & \textit{Tun} & 0 & 0 & 2 & \cellcolor{verygood}18*** & \cellcolor{good}41* & 0 & 4 & 8 & 33 & 19 & 0 & 0 & 0 & \cellcolor{excellent}52*** & \cellcolor{excellent}91*** \\
  \midrule

  \multirow{2}{*}{\textbf{H3}} & \textit{Std} & 0 & 0 & 1 & \cellcolor{good}6*** & \cellcolor{neutral}0*** & 0 & \cellcolor{good}6* & 5 & 22 & \cellcolor{verygood}25** & 0 & \cellcolor{neutral}0** & \cellcolor{good}10** & \cellcolor{excellent}60*** & \cellcolor{excellent}97*** \\
  & \textit{Tun} & 0 & 0 & \cellcolor{neutral}0*** & \cellcolor{good}8*** & \cellcolor{verygood}10*** & 0 & 5 & 7 & \cellcolor{verygood}40*** & 38 & 0 & \cellcolor{neutral}0** & \cellcolor{good}10** & \cellcolor{excellent}60*** & \cellcolor{excellent}97*** \\
  \midrule

  \multirow{2}{*}{\textbf{H4}} & \textit{Std} & \cellcolor{neutral}0*** & \cellcolor{neutral}0*** & \cellcolor{neutral}0*** & \cellcolor{neutral}0*** & \cellcolor{neutral}0*** & 0 & \cellcolor{good}9** & 6 & \cellcolor{good}23* & \cellcolor{verygood}18*** & 0 & 0 & \cellcolor{good}2* & \cellcolor{good}8*** & \cellcolor{neutral}0*** \\
  & \textit{Tun} & \cellcolor{neutral}0*** & \cellcolor{neutral}0*** & \cellcolor{neutral}0*** & \cellcolor{neutral}0*** & \cellcolor{good}9*** & 0 & \cellcolor{good}9** & 6 & \cellcolor{verygood}12*** & \cellcolor{verygood}26*** & 0 & 0 & \cellcolor{good}10** & \cellcolor{verygood}46* & \cellcolor{verygood}15*** \\
  \midrule

  \multirow{2}{*}{\textbf{H5}} & \textit{Std} & \cellcolor{neutral}0** & \cellcolor{neutral}0*** & \cellcolor{neutral}0*** & \cellcolor{good}15* & 35 & 0 & \cellcolor{good}8* & 6 & \cellcolor{verygood}35** & \cellcolor{verygood}14*** & 0 & 0 & \cellcolor{good}10** & \cellcolor{excellent}63*** & \cellcolor{excellent}91*** \\
  & \textit{Tun} & \cellcolor{neutral}0** & 0 & \cellcolor{neutral}0*** & 18 & \cellcolor{good}42* & 0 & \cellcolor{good}5* & 6 & \cellcolor{verygood}38** & \cellcolor{verygood}18** & 0 & 0 & \cellcolor{good}10** & \cellcolor{excellent}63*** & \cellcolor{excellent}94*** \\
  \midrule

  \multirow{2}{*}{\textbf{H6}} & \textit{Std} & \cellcolor{neutral}0** & \cellcolor{neutral}0*** & \cellcolor{neutral}0*** & \cellcolor{verygood}11*** & 33 & 0 & \cellcolor{good}9** & \cellcolor{good}8* & 29 & \cellcolor{verygood}10*** & 0 & 0 & \cellcolor{good}10** & \cellcolor{excellent}63*** & \cellcolor{excellent}86*** \\
  & \textit{Tun} & \cellcolor{neutral}0** & \cellcolor{neutral}0*** & \cellcolor{neutral}0*** & 19 & 38 & 0 & \cellcolor{good}7* & 7 & \cellcolor{verygood}43*** & 24 & 0 & 0 & \cellcolor{good}10** & \cellcolor{excellent}63*** & \cellcolor{excellent}97*** \\
  \midrule

  \multirow{2}{*}{\textbf{H7}} & \textit{Std} & 0 & 0 & 0 & \cellcolor{verygood}18*** & \cellcolor{verygood}39*** & 0 & 4 & 7 & 29 & \cellcolor{good}17* & 0 & 0 & \cellcolor{good}10** & \cellcolor{excellent}63*** & \cellcolor{excellent}94*** \\
  & \textit{Tun} & 0 & 0 & 0 & \cellcolor{verygood}17*** & \cellcolor{verygood}37*** & 0 & 4 & \cellcolor{good}4* & 22 & \cellcolor{verygood}15** & 0 & 0 & \cellcolor{good}10** & \cellcolor{excellent}63*** & \cellcolor{excellent}92*** \\
  \midrule

  \multirow{2}{*}{\textbf{H8}} & \textit{Std} & 0 & 0 & 6 & \cellcolor{good}8** & 30 & 0 & \cellcolor{good}8* & 7 & 26 & \cellcolor{verygood}10*** & 0 & \cellcolor{neutral}0** & 10 & 31 & \cellcolor{good}2*** \\
  & \textit{Tun} & 0 & 0 & \cellcolor{good}10** & \cellcolor{verygood}23*** & \cellcolor{good}29* & 0 & 6 & 6 & \cellcolor{verygood}40** & 28 & 0 & 0 & 10 & 40 & \cellcolor{verygood}16*** \\
  \midrule

  \multirow{2}{*}{\textbf{H9}} & \textit{Std} & 0 & 0 & \cellcolor{good}6* & 12 & \cellcolor{good}31* & 0 & \cellcolor{good}7* & \cellcolor{good}8* & 23 & \cellcolor{verygood}12*** & 0 & 0 & \cellcolor{good}10** & 44 & \cellcolor{verygood}32*** \\
  & \textit{Tun} & 0 & 0 & \cellcolor{good}7* & 7 & 24 & 0 & 4 & \cellcolor{good}7* & \cellcolor{good}30* & \cellcolor{good}25* & 0 & 0 & \cellcolor{good}10** & \cellcolor{excellent}60*** & \cellcolor{excellent}73*** \\
  \midrule

  \multirow{2}{*}{\textbf{H10}} & \textit{Std} & 0 & 0 & \cellcolor{good}6* & 9 & 36 & 0 & \cellcolor{good}8* & 8 & \cellcolor{verygood}41** & \cellcolor{verygood}15*** & 0 & \cellcolor{neutral}0** & 10 & 43 & \cellcolor{verygood}11*** \\
  & \textit{Tun} & 0 & 0 & \cellcolor{good}8** & 13 & 29 & 0 & 1 & 6 & 29 & 18 & 0 & \cellcolor{neutral}0** & 10 & \cellcolor{verygood}46** & \cellcolor{good}1*** \\
  \bottomrule[1.5pt]

  \end{tabular}
  \end{adjustbox}
  
  \vspace{0.2em}
  \parbox{0.95\textwidth}{\small
  \noindent\textbf{Note:} \textit{Std} refers to using the parameter configuration determined by \texttt{irace-evo}, while 
  \textit{Tun} uses parameters optimized specifically for each dataset (B1, B2, B3).
  Higher win rates indicate stronger generalization or dataset-specific adaptation.
  }

\end{table}

\paragraph{Cost.} We used Claude~Haiku~3.5 to show that \texttt{irace-evo} can surpass state-of-the-art performance without requiring a more expensive and computationally demanding LLM like GPT-4, while remaining fully compatible with models such as those from OpenAI. The total cost for all ten \texttt{irace-evo} runs---each consisting of five iterations---was only \euro1.57, averaging less than \euro0.20 per \texttt{irace-evo} run. This is an exceptionally low expense given the large volume of input and output tokens processed, highlighting the practical efficiency of our approach.

\begin{table}[!t]
  \centering
  \renewcommand{\arraystretch}{1.15}
  \setlength{\tabcolsep}{3pt}
  \small

  \caption{Cases in which some of the produced algorithm variants produce new best solutions (n$\geq$500)}
  \label{tab:improvements}

  \begin{adjustbox}{max width=\textwidth}
  \begin{tabular}{p{1.5cm}ccccp{4.5cm}p{4.5cm}}
    \toprule[1.3pt]
    \rowcolor{headerback}
    \textbf{Dataset} & \textbf{$n$} & \textbf{\#} & \textbf{Old Best} & \textbf{New Best} & \textbf{Best H (Std)} & \textbf{Best H (Tuned)} \\
    \midrule[1pt]

    \multirow{5}{*}{\textbf{B1}} 
    & 500 & 1 & \cellcolor{tableback}61760 & \cellcolor{accentLighter}\textbf{61750} & \cellcolor{tableback}H2-H3, H8-H10 & \cellcolor{tableback}H2, H8-H10 \\
    \cmidrule{2-7}
    & 1000 & 6 & \cellcolor{tableback}125520 & \cellcolor{accentLighter}\textbf{125510} & \cellcolor{tableback}H1-H2, H9 & \cellcolor{tableback}H1-H2, H5-H8, H10 \\
    \cmidrule{2-7}
    & \multirow{3}{*}{2000} 
    & 5 & \cellcolor{tableback}245060 & \cellcolor{accentLighter}\textbf{245050} & \cellcolor{tableback}H2, H5-H10 & \cellcolor{tableback}H1-H2, H5-H7, H9-H10 \\
    & & 8 & \cellcolor{tableback}256970 & \cellcolor{accentLighter}\textbf{256950} & \cellcolor{tableback}H1-H10 & \cellcolor{tableback}H1-H10 \\
    & & 9 & \cellcolor{tableback}258450 & \cellcolor{accentLighter}\textbf{258430} & \cellcolor{tableback}H1-H10 & \cellcolor{tableback}H1-H2, H5-H10 \\

    \midrule

    \multirow{6}{*}{\textbf{B2}} 
    & \multirow{2}{*}{1000} & 9 & \cellcolor{tableback}78369 & \cellcolor{accentLighter}\textbf{78365} & \cellcolor{tableback}H1, H5, H7, H9 & \cellcolor{tableback}H3-H6 \\
    & & 10 & \cellcolor{tableback}84503 & \cellcolor{accentLighter}\textbf{84502} & \cellcolor{tableback}H1-H8, H10 & \cellcolor{tableback}H1-H6, H8-H10 \\
    \cmidrule{2-7}
    & \multirow{4}{*}{2000} & 3 & \cellcolor{tableback}161857 & \cellcolor{accentLighter}\textbf{161855} & \cellcolor{tableback}H4 & \cellcolor{tableback}H3 \\
    & & 4 & \cellcolor{tableback}161064 & \cellcolor{accentLighter}\textbf{161063} & \cellcolor{tableback}H2, H7 & \cellcolor{tableback}H1-H3, H5-H6, H8-H9 \\
    & & 5 & \cellcolor{tableback}164611 & \cellcolor{accentLighter}\textbf{164602} & \cellcolor{tableback}--- & \cellcolor{tableback}H3-H4 \\
    & & 10 & \cellcolor{tableback}153256 & \cellcolor{accentLighter}\textbf{153255} & \cellcolor{tableback}H6-H7, H10 & \cellcolor{tableback}H2, H5-H8 \\

    \midrule

    \multirow{15}{*}{\textbf{B3}} 
    & 500 & 10 & \cellcolor{tableback}90543 & \cellcolor{accentLighter}\textbf{90541} & \cellcolor{tableback}H1-H10 & \cellcolor{tableback}H1, H3-H10 \\
    \cmidrule{2-7}
    & \multirow{4}{*}{1000} & 1 & \cellcolor{tableback}176950 & \cellcolor{accentLighter}\textbf{176948} & \cellcolor{tableback}H1-H3, H5-H7 & \cellcolor{tableback}H1-H3, H4-H7, H9-H10 \\
    & & 2 & \cellcolor{tableback}180989 & \cellcolor{accentLighter}\textbf{180987} & \cellcolor{tableback}H1, H5-H7, H9 & \cellcolor{tableback}H1, H4-H7 \\
    & & 7 & \cellcolor{tableback}178183 & \cellcolor{accentLighter}\textbf{178181} & \cellcolor{tableback}H1-H3, H5-H7, H9-H10 & \cellcolor{tableback}H1, H4-H10 \\
    & & 10 & \cellcolor{tableback}176910 & \cellcolor{accentLighter}\textbf{176902} & \cellcolor{tableback}H1-H3, H5-H7, H9-H10 & \cellcolor{tableback}H1-H3, H5-H9 \\
    \cmidrule{2-7}
    & \multirow{10}{*}{2000} & 1 & \cellcolor{tableback}356235 & \cellcolor{accentLighter}\textbf{356212} & \cellcolor{tableback}H2-H3, H5, H7 & \cellcolor{tableback}H1-H3, H5-H7, H9 \\
    & & 3 & \cellcolor{tableback}364543 & \cellcolor{accentLighter}\textbf{364513} & \cellcolor{tableback}H1-H3, H5-H7, H9 & \cellcolor{tableback}H1-H3, H5-H7, H9 \\
    & & 4 & \cellcolor{tableback}356960 & \cellcolor{accentLighter}\textbf{356958} & \cellcolor{tableback}H1-H3, H5-H7 & \cellcolor{tableback}H1-H3, H5-H7 \\
    & & 5 & \cellcolor{tableback}365564 & \cellcolor{accentLighter}\textbf{365527} & \cellcolor{tableback}H1-H3, H5, H7 & \cellcolor{tableback}H1-H3, H5-H7 \\
    & & 6 & \cellcolor{tableback}365116 & \cellcolor{accentLighter}\textbf{365110} & \cellcolor{tableback}H1-H3, H5-H6 & \cellcolor{tableback}H1, H3, H5-H7 \\
    & & 7 & \cellcolor{tableback}360830 & \cellcolor{accentLighter}\textbf{360792} & \cellcolor{tableback}H1, H3, H5-H7 & \cellcolor{tableback}H1, H3, H5-H7 \\
    & & 8 & \cellcolor{tableback}371776 & \cellcolor{accentLighter}\textbf{371767} & \cellcolor{tableback}H1-H3, H5-H7 & \cellcolor{tableback}H1, H3, H5-H7 \\
    & & 9 & \cellcolor{tableback}355733 & \cellcolor{accentLighter}\textbf{355691} & \cellcolor{tableback}H1, H5-H7 & \cellcolor{tableback}H1, H3, H5-H7 \\
    & & 10 & \cellcolor{tableback}357039 & \cellcolor{accentLighter}\textbf{357030} & \cellcolor{tableback}H1-H3, H5-H7 & \cellcolor{tableback}H1-H3, H5-H7, H9 \\

    \bottomrule[1.3pt]
  \end{tabular}
  \end{adjustbox}

  \vspace{0.2em}
  \parbox{0.95\textwidth}{\small
  \noindent\textbf{Note:} This table shows only instances where at least one of the produced algorithm variants achieves a better result than the ``Old Best'' value. \textit{Old Best} refers to the best result from the original algorithm, while \textit{New Best} shows the best value among all produced algorithm variants. Again, the analysis is separately shown for \textit{Std} and \textit{Tuned}. For space limitations, only instances with size $\geq$ 500 are included.
  }

\end{table}
\paragraph{Code Analysis.}
Among the algorithm variants generated by \texttt{irace-evo}, two stand out for their strong and consistent performance across all datasets (B1–B3): $H_5$ and $H_7$. An analysis of their LLM-generated placement heuristics reveals that both extend the original placement heuristic (Section~\ref{subsubsec:original-heuristic}) by incorporating adaptive terms that balance cost, load utilization, and search context. In particular, $H_5$ features a \emph{dynamic efficiency score} that combines cost efficiency, bin utilization, and the number of remaining items. This formulation penalizes underutilized bins and adjusts the evaluation according to the remaining search space, encouraging more balanced packing decisions and better scalability on large instances.

\begin{lstlisting}[style=mintedclone2]
double evaluate_placement_quality(...) {
    double utilization_factor = 
        1.0 - (double(new_load) / bin_capacities[new_bin_type]);
    double cost_efficiency = bin_costs[new_bin_type] / (new_load + 1.0);
    double remaining_factor = 
        (remaining_items > 0) ? 1.0 / remaining_items : 1.0;
    return cost_efficiency * (1.0 + utilization_factor) * (1.0 + remaining_factor);
}
\end{lstlisting}

On the other side, $H_7$ makes use of an \emph{enhanced cost-efficiency metric} with a utilization bias. By introducing a \texttt{remaining\_pressure} factor (see the code below), it adapts the decision rule along the solution construction, maintaining efficiency when few items remain and preventing premature convergence toward suboptimal bin types.

\begin{lstlisting}[style=mintedclone2]
double evaluate_placement_quality(...) {
    double base_ratio = bin_costs[new_bin_type] / (new_load + 1e-8);
    double utilization_factor = 
        1.0 - (new_load / (bin_capacities[new_bin_type] + 1e-8));
    double remaining_pressure = 
        static_cast<double>(remaining_items) / (num_items + 1e-8);
    return base_ratio * (1.0 + utilization_factor * remaining_pressure);
}
\end{lstlisting}

Overall, both $H_5$ and $H_7$ enrich the original placement heuristic by embedding context-aware dynamics. While $H_5$ prioritizes balance between cost and utilization efficiency, $H_7$ introduces adaptive pressure mechanisms that adjust sensitivity to remaining items, leading to better placement decisions across instances with diverse characteristics.


\section{Discussion and Conclusions}

In this work, we introduced the first version of an extension to \texttt{irace} called \texttt{irace-evo}, which adds the ability to evolve code (e.g., algorithmic components). The results show that irace-evo was able to discover---not just one, but several---item placement heuristics that outperformed the state-of-the-art for the Variable-Sized Bin Packing Problem when being used within the Construct, Merge, Solve \& Adapt algorithm implemented in C++. We showed that \texttt{irace-evo} can identify these placement heuristics together with well-working parameter configurations at very low cost, using Claude Haiku 3.5 for all experiments with a total expense below \euro 2.

Our work is an example of recent research in optimization that shows a shift toward systems in which code adapts dynamically to changing conditions. Leveraging evolutionary computation and LLM-based agent frameworks, the ability to autonomously detect shifts and restructure itself is set to define next-generation software~\cite{zhang2025darwingodelmachineopenended,wang2025huxleygodelmachinehumanlevelcoding}.

In our study, we applied the Always-From-Original principle to increase exploration and shorten evolutionary chains. Using a low-cost LLM (Claude Haiku 3.5), we observed a clear improvement by setting \texttt{top\_p} to 0.9, restricting sampling to tokens covering 90\% of the probability mass. This highlights that hyperparameters strongly affect code-generation quality: temperature alone is insufficient, and other sampling parameters require careful tuning. Both considerations should be taken into account in fully evolutionary system designs~\cite{10.1145/3731567}.

Future work includes the extension to the simultaneous improvement of more than one algorithmic component. Moreover, we envisage multi-code evolution to different parts of the algorithm, extending beyond heuristic refinement to produce code optimized for specific hardware. For further details and experimental results, please refer to the supplementary material which also includes an example of a JSON file for the configuration of \texttt{irace-evo}.

\subsubsection{Acknowledgements} The research presented in this paper was supported by grant PID2022-136787NB-I00 funded by MCIN/AEI/10.13039/501100011033 (Christian Blum) and is part of the R\&D project PID2022-138283NB-I00, funded by MICIU/AEI/10.13039/501100011033 and “FEDER --- A way of making Europe” (Camilo Chacón Sartori).

\bibliographystyle{splncs04}  
\bibliography{ref}  

\newpage
\appendix

\section{Appendix A --- Additional Analyses}
\subsection{Error Rate in Code Generation}

Each algorithmic variant produced by the LLM may fail to compile when using a compiled language such as C++. To mitigate this, \texttt{irace-evo} includes a compilation–retry mechanism: whenever a variant fails to compile, the system attempts recompilation up to $n$ times. This parameter, along with the choice of compiler and its flags, can be configured in the \texttt{CodeEvolution.json} file.

For the Variable-Sized Bin Packing Problem (VSBPP) solved with the Construct, Merge, Solve \& Adapt (CMSA) framework, the observed compilation error rate is reported in Table~\ref{tab:error}.

Our experiments show that compilation errors primarily arise from prompt design, the selected LLM, and its hyperparameters. Prompt specificity plays a central role: if the prompt is underspecified, the code space available to the model becomes too broad, increasing the likelihood that the LLM will drift from the intended structure, generate a degraded variant, or even violate the Always-From-Original principle. Model size also matters. Small Language Models (SLMs) tend to produce more compilation errors because their pretrained knowledge is more limited or less aligned with the target problem. Hyperparameters such as temperature and top-$p$ further influence generation quality. In particular, we observe that top-$p$ plays a crucial role in improving response quality; for instance, with Claude Haiku 3.5, lowering top-$p$ from 1.0 to 0.9 results in a marked increase in the quality of the generated code variants.

\begin{table}[h]
\centering
\begin{tabular}{c|c|c|c|c|c}
\hline
\textbf{CMSA} & \textbf{Compilation Errors} & \textbf{Total Iterations} & \textbf{Total Variants} & \textbf{Error Rate (\%)} \\
\hline
1  & 4  & 6 & 30 & 13.33 \\
2  & 6  & 6 & 30 & 20.00 \\
3  & 2  & 6 & 30 & 6.66  \\
4  & 0  & 5 & 25 & 0.00  \\
5  & 0  & 5 & 25 & 0.00  \\
6  & 4  & 8 & 40 & 10.00 \\
7  & 2  & 5 & 25 & 8.00  \\
8  & 2  & 8 & 40 & 5.00  \\
9  & 1  & 7 & 35 & 2.85  \\
10 & 0  & 7 & 35 & 0.00  \\
\hline
\textbf{Total} & \textbf{21} & \textbf{63} & \textbf{315} & \textbf{6.67} \\
\hline
\end{tabular}
\caption{Summary of compilation errors observed across CMSA runs during the \texttt{irace-evo} process.}\label{tab:error}
\end{table}

\subsection{Heuristics With and Without Parameter Tuning}

\begin{figure}
\centering
  \includegraphics[width=1\linewidth]{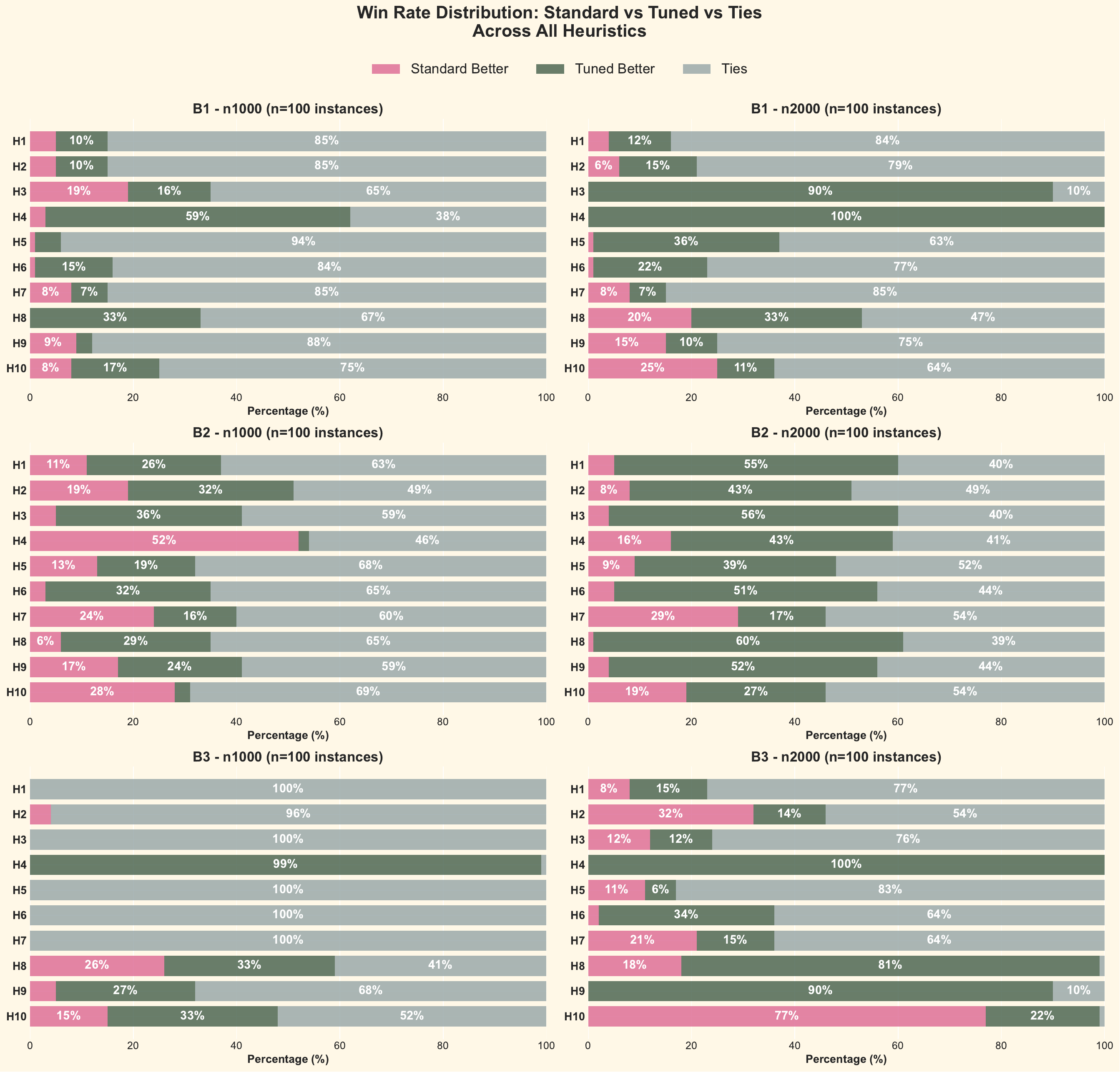}
  \caption{Performance comparison of heuristics with and without independent tuning of parameters.}

  \label{fig:win-tuning-standard}
\end{figure}
To evaluate the effectiveness of the parameter configurations generated by \texttt{irace-evo}, we conducted an additional tuning phase using \texttt{irace} with a larger budget (\texttt{maxExperiments = 2000}). The ten heuristics produced by \texttt{irace-evo} for the VSBPP---originally obtained under a much smaller budget of \texttt{maxExperiments = 300}---were used as the starting point.

The results indicate that many heuristics benefit from further tuning, while in some instances the improvements are negligible. This suggests that \texttt{irace-evo}, even with a minimal budget, is already capable of generating parameter configurations that, although not necessarily optimal, are strong and competitive.

A comparison between heuristics with and without parameter tuning is shown in Figure~\ref{fig:win-tuning-standard}. A notable case occurs with the B2 dataset at $n=1000$, where in many instances the untuned heuristic outperforms the tuned one. Conversely, for the B3 dataset at $n=1000$, there is no significant difference between the tuned and untuned heuristics, as several ties are observed.

A direction for future work is to increase \texttt{maxExperiments} in \texttt{irace-evo} runs to assess whether further improvements can be achieved, or if simply increasing the number of \texttt{codeEvolutionVariants} is sufficient. Given that our Always-From-Original principle is designed for ``short evolutions,'' raising the budget may not produce substantial differences unless the dynamic prompts are made more sophisticated.

\section{Appendix B --- \texttt{code-evolution.json}}

For \texttt{irace-evo} to operate, the \texttt{code-evolution.json} file is required, as specified in the \texttt{irace} configuration file \texttt{scenario.txt}.

Below, we present the file used in our experiments. Note that CMSA is an algorithm that internally uses CPLEX, so this file allows configuring the C++ compiler flags. While using Python would make the setup simpler, this \texttt{irace-evo} extension demonstrates that it is possible to evolve complex C++ code---a programming language that offers substantial flexibility in its configuration.

\begin{lstlisting}
{
  "_comment": "irace-evo Configuration for VSBPP-CMSA Evolution",
  "_version": "1.0",

"problem_context": {
  "problem_name": "Variable-sized bin packing problem (VSBPP)",
  "problem_description": "Given a set of items with varying sizes and a set of bins with different capacities and costs, the goal is to assign all items to bins while minimizing the total cost or number of bins used. This is a combinatorial optimization problem focusing on efficient packing and cost-effective bin utilization. Any heuristic should first extract the most relevant constraints and features from the problem context and then propose a solution that is both novel and minimal in structure, similar in size and complexity to the original heuristic.",
  "algorithm_approach": "Construct, Merge, Solve And Adapt (CMSA) metaheuristic: candidate solutions are iteratively constructed and merged, then optimized via an exact solver or heuristic adaptation. Solution components are evaluated and adapted based on their contribution to overall packing efficiency and feasibility. Heuristics should be concise and maintain a minimal structure, avoiding unnecessary expansions while introducing novel ideas grounded in the problem context.",
  "optimization_objective": "Minimize the total number of bins used or total cost while ensuring all items are packed without exceeding bin capacities. Heuristics should remain compact and interpretable, with a structure comparable to the original, while meaningfully improving decision-making based on the problem features.",
  "key_challenges": [
    "Designing heuristics that are effective, novel, and minimal in structure, maintaining a size and complexity similar to existing simple heuristics.",
    "Merging solution components to explore high-quality and unconventional combinations without inflating heuristic complexity.",
    "Integrating solver-based adaptations efficiently, avoiding overly complex formulas or arbitrary calculations.",
    "Handling heterogeneous bin capacities and item sizes by identifying problem-specific features that inspire concise yet original heuristic rules.",
    "Preventing premature convergence to suboptimal bin assignments by encouraging diverse but structurally minimal heuristics.",
    "Balancing exploration of new packing configurations with exploitation of high-quality solutions, while keeping heuristics compact and interpretable.",
    "Ensuring heuristics are linear where applicable, derived from problem features, and not from ad-hoc tricks or magic numbers.",
    "Avoiding hard-coded thresholds or unjustified shortcuts that could increase complexity without real benefit.",
    "Eliminating mathematical tricks that are ungrounded in the problem context.",
    "Guaranteeing that heuristics generalize well across different instance sizes and distributions while remaining minimal."
  ],
  "performance_considerations": "Evaluating candidate packings is computationally demanding. Efficiency requires incremental feasibility checks, optimized merging strategies, and parallelization wherever feasible. Any heuristic should remain minimal in structure to preserve computational efficiency and interpretability.",
  "domain_knowledge": "Relevant heuristics for VSBPP include prioritizing large items first, considering bin cost-effectiveness, leveraging lower/upper bound approximations, and exploring combinations of bin sizes and item distributions. New heuristics should be compact, minimally structured, and aligned with the problem context while introducing novel decision rules."
},


  "language_config": {
    "language": "cpp",
    "comment": "C++ implementation of CMSA for VSBPP problem with packing-based heuristics"
  },

  "source_config": {
    "source_file": "./src/cmsa_set_covering.cpp",
    "function_name": "evaluate_placement_quality",
    "function_signature": "double evaluate_placement_quality(int current_bin_type, int new_bin_type, int current_load, int item_weight, int item_index, const vector<int>& bin_costs, const vector<int>& bin_capacities, const vector<int>& item_weights,int num_items,int num_bin_types,int remaining_items)",
    "includes": [
      "<vector>",
      "<algorithm>", 
      "<random>",
      "<iostream>",
      "<fstream>",
      "<cmath>",
      "<list>",
      "<set>"
    ],
    "dependencies": [],
    "comment": ""
  },

  "build_config": {
    "cpp": {
      "compiler": "g++",
      "flags": [
        "-m64",
        "-O",
        "-fPIC",
        "-fexceptions",
        "-DNDEBUG",
        "-DIL_STD",
        "-std=c++17",
        "-fpermissive",
        "-w",
        "-I/home/user/CPLEX_Studio221/cplex/include",
        "-I/home/user/CPLEX_Studio221/concert/include"
      ],
      "link_flags": [
        "-L/home/user/CPLEX_Studio221/cplex/lib/x86-64_linux/static_pic",
        "-lilocplex",
        "-lcplex",
        "-L/home/user/CPLEX_Studio221/concert/lib/x86-64_linux/static_pic",
        "-lconcert",
        "-lm",
        "-pthread",
        "-lpthread",
        "-ldl"
      ],
      "include_paths": [
        "./src"
      ],
      "library_paths": [
        "/home/user/CPLEX_Studio221/cplex/lib/x86-64_linux/static_pic",
        "/home/user/CPLEX_Studio221/concert/lib/x86-64_linux/static_pic"
      ],
      "libraries": [
        "ilocplex",
        "cplex",
        "concert",
        "m",
        "pthread",
        "dl"
      ],
      "output_dir": "./bin",
      "compile_timeout": 30
    }
  },

  "llm_config": {
    "api_provider": "anthropic",
    "model": "claude-3-5-haiku-latest",
    "api_key": "",
    "max_retries": 3,
    "temperature": 1,
    "use_dynamic_prompting": true,
    "max_tokens": 2000,
    "timeout": 60
  },

  "progressive_context": {
    "enabled": true,
    "first_iteration_full_context": true,
    "reduction_schedule": [1.0, 0.7, 0.5, 0.3, 0.3],
    "min_context_ratio": 0.2,

    "selection_strategies": {
      "performance_based": true,
      "function_frequency": true,
      "dynamic_prompting_focus": true,
      "parameter_correlation": true
    }
  },
  
  "evolution_config": {
    
    "max_compilation_failures": 3,
    "intelligent_error_correction": true,
    "max_error_correction_attempts": 2,

    "available_strategies": [
      "innovate_heuristic_design"
    ],

    "strategy_selection": "weighted",
    "strategy_weights": {
      "innovate_heuristic_design": 1
    },

    "comment": "Custom VSBPP strategy focuses on domain-specific improvements for CMSA packing solutions"
  }
}

\end{lstlisting}

\end{document}